\documentclass[10pt,twocolumn]{IEEEtran}

\usepackage{color}
\usepackage{makecell}
\usepackage{xcolor}
\usepackage{soul}
\usepackage{amsmath}
\usepackage{graphics}
\usepackage{setspace}
\usepackage{cite}
\usepackage{latexsym}
\usepackage{float}
\usepackage{epsfig}
\usepackage{multirow}
\usepackage{cite,cases,url}
\usepackage{amssymb}
\usepackage{graphicx}
\usepackage{epstopdf}
\usepackage{balance}
\usepackage{subcaption}
\usepackage{enumerate}
\usepackage{array}
\usepackage{algorithmic}
\usepackage{algorithm}
\usepackage{enumitem}
\usepackage{dblfloatfix}
\usepackage{tabularx}
\usepackage{siunitx}
\usepackage[normalem]{ulem}
\useunder{\uline}{\ul}{}
\newcolumntype{Y}{>{\centering\arraybackslash}X}

\newcolumntype{L}{>{\centering\arraybackslash}m{5cm}}
\newcolumntype{K}{>{\centering\arraybackslash}m{6cm}}
\newcolumntype{P}{>{\centering\arraybackslash}m{2.3cm}}
\newcolumntype{M}{>{\raggedright\arraybackslash}m{2cm}}
\newcolumntype{N}{>{\raggedright\arraybackslash}m{2.5cm}}

\begin{document}

\title{{Handover Experiments with UAVs: Software Radio \\Tools and Experimental Research Platform}}

\author{
\IEEEauthorblockN{Keith Powell, Andrew Yingst, Talha Faizur Rahman, and Vuk Marojevic
}
}

%
\maketitle
\begin{abstract}

Mobility management is the key feature of cellular networks. When integrating unmanned aerial vehicles (UAVs) into cellular networks, their cell association needs to be carefully managed for coexistence with other cellular users. UAVs move in three dimensions and may traverse several cells on their flight path, and so may be subject to several handovers. In order to enable research on mobility management with UAV users, this paper describes the design, implementation, and testing methodology for handover experiments with aerial users. We leverage software-defined radios (SDRs) and implement a series of tools for preparing the experiment in the laboratory and for taking it outdoors for field testing. We use solely commercial off-the-shelf hardware, open-source software, and an experimental license to enable reproducible and scalable experiments. Our initial outdoor results with two SDR base stations connected to an open-source software core network, implementing the 4G long-term evolution protocol, and one low altitude UAV user equipment demonstrate the handover process.

\end{abstract}



\IEEEpeerreviewmaketitle

\section{Introduction}
\label{sec:intro}
Ever since the idea of the software radio was conceived, the development of reprogrammable radio technology with more advanced innovations has been desired. 
Software-defined radio (SDR) platforms enable cognitive radio and dynamic spectrum access \cite{mitola_cr} because of their 
flexibility and cost efficiency to drive the communications protocols forward. Several open-source SDR projects have developed hardware and software to enable research and development in wireless communications and networking. 

A few open-source cellular radio stacks exist supporting end-to-end networks for use with SDRs. Among them is srsLTE (later renamed as srsRAN) \cite{gomez2016srslte}, which supports the full 4G long-term evolution (LTE) network end-to-end with many advanced LTE functionalities, such as handovers, multiple input, multiple output (MIMO), and carrier aggregation. Additionally, 5G non-standalone mode on their user equipment (UE) is currently available with 5G network support expected in future srsRAN releases.  

Next generation wireless networks will integrate a wide range of emerging technologies, including unmanned aerial vehicles (UAVs) and intelligent reflective surfaces. 
It is, therefore, essential to do feasibility studies of these enabling technologies with current technologies. Of particular interest are UAVs which are being investigated by researchers and engineers due to their flexibility in service provisioning \cite{8682048, 8436792}.

Garcia et al. \cite{UAV_SDR} discuss an experimental setup using SDRs to determine their ability to communicate with aerial nodes, UAVs and other aircraft. An Ettus Research Universal  Software  Radio  Peripheral (USRP) running srsRAN is lifted by a UAV to perform throughput measurements between the aerial node and the ground node. Directional antennas are used on the ground node and pointed toward the sky where the flight path of the UAV is focused. 

Reference \cite{SDR_Airforce} describes laboratory and field trials for performing handover to measure the effectiveness of their solution that compensates for the Doppler shift. For both sets of experiments, two LTE base station (eNodeB or eNB) are set up with an LTE UE. The laboratory tests use SDRs for implementing channel effects, such as path loss and Doppler shifts that may be seen in an over-the-air experiment. For the field tests, the UE is attached to a car and driven between the base stations (BSs) to record real channel effects.

UAVs provide unique mobility features and cellular networks are uniquely qualified in providing mobility management. But there is a lack of large-scale experimental research opportunities with UAVs and advanced wireless protocols.  
The Aerial Experimentation and Research Platform for Advanced Wireless (AERPAW) closes this gap and provides the framework and at-scale outdoor testbed for 5G research with UAVs \cite{aerpaw_vtc}. This paper describes the design, implementation, and testing methodology for enabling and evaluating handover experiments with aerial users. In an effort to enable reproducible and scalable experiments, we leverage SDRs and implement a series of tools for preparing the experiment in the laboratory and for taking it outdoors for field testing. We use solely commercial off-the-shelf hardware, open source software, and an experimental license for radiation of broadband multicarrier signals. Our initial outdoor results with two SDR BSs and an open-source core network (CN), implementing the 4G LTE network, and a low altitude small UAV that carries an SDR implementing the 4G UE demonstrate the handover process. 

The rest of the paper is organized as follows: Section 2 discusses the experiment development and optimization flow by defining a series of benchmarks. The benchmarks give methods for determining the effectiveness of experiments and processes for carrying them out. Section 3 summarizes the use of ZeroMQ (ZMQ) with srsRAN, and its use with GNU Radio for creating advanced LTE experiments without SDR and radio frequency (RF) hardware. Section 4 examines the requirements for setting up a real world outdoor SDR experiment using S1 handover with UAVs and our measurement results. The transitioning from laboratory to outdoor experiments is discussed in terms of hardware requirements, backhaul, and additional considerations. The paper concludes with expectations of future experiments using the proposed platform.


\section{Experiment Optimization}
\label{sec:benchmark}
As a result of 
the highly-configurable nature of SDRs, it can be difficult to make ideal hardware selections and optimize software parameters in order to achieve maximum performance from the radios. The process of benchmarking the radio performance can be impacted by the physical environment, SDR and RF front end selection, transmitted waveform, SDR software parameters, UAV payload limitations, and spectrum availability. We show the process using srsRAN and USRPs, but the same process is generally applicable to other SDR experiments. This section discusses various benchmarks used to define performance and our methodology for maximizing performance for a given scenario.

\subsection{Benchmarks}

Benchmarks are essential for measuring and comparing performance and for fine tuning parameters to achieve the desired or 
the best possible 
outcomes of an experiment. Different benchmarks may be more beneficial than others based on the type of experiment.  Additionally, running benchmarks in parallel can be effective for improving the overall performance and understanding of an experiment.

\begin{figure}[h]
    \begin{subfigure}{0.47\textwidth}
      \includegraphics[width=\textwidth]{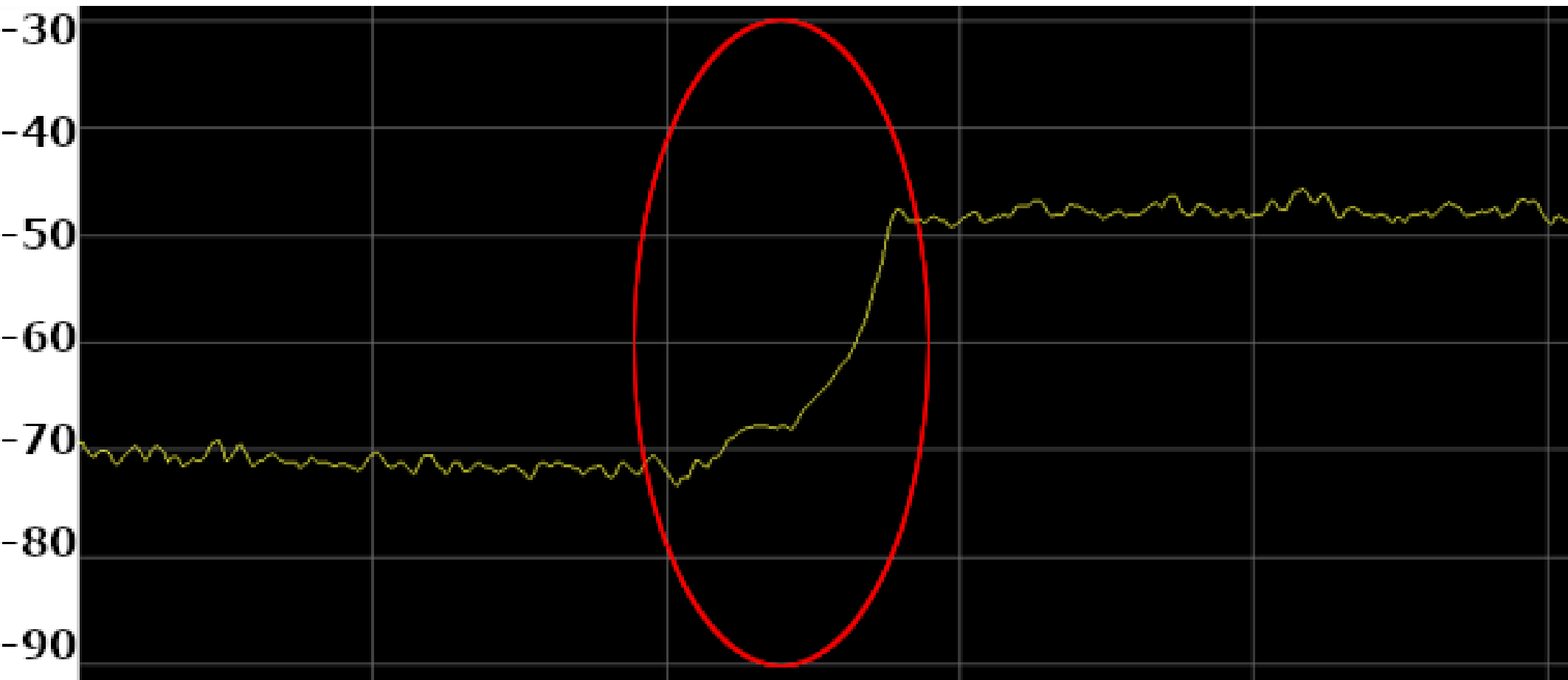}
      \caption{Transmit gain 73 with minimal distortion.}
    \end{subfigure}
    \begin{subfigure}{0.47\textwidth}
      \includegraphics[width=\textwidth]{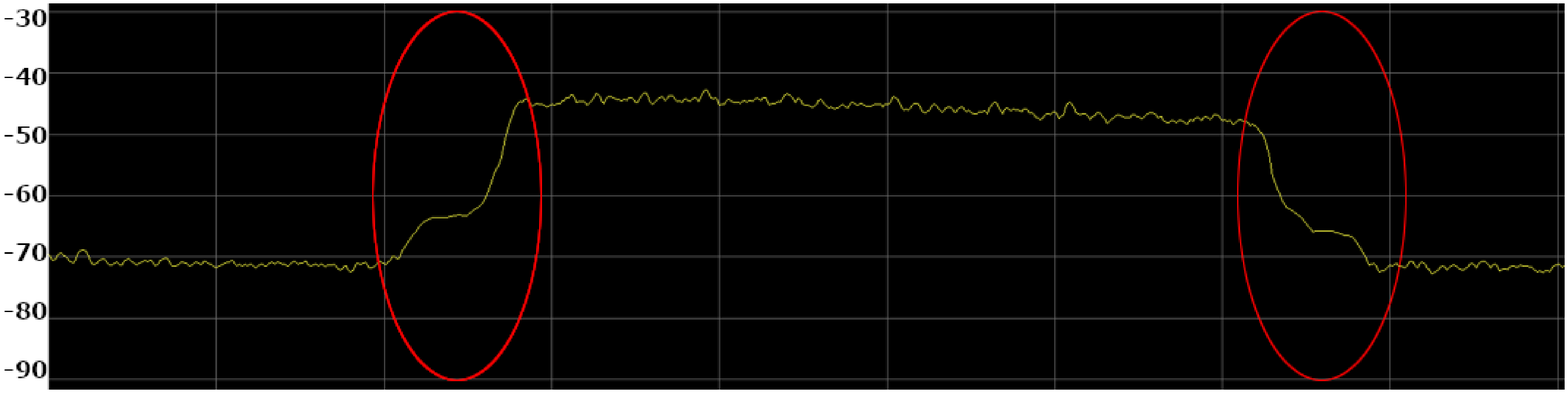}
      \caption{Transmit gain 76 with mild distortion.}
    \end{subfigure}
    \begin{subfigure}{0.47\textwidth}
      \includegraphics[width=\textwidth]{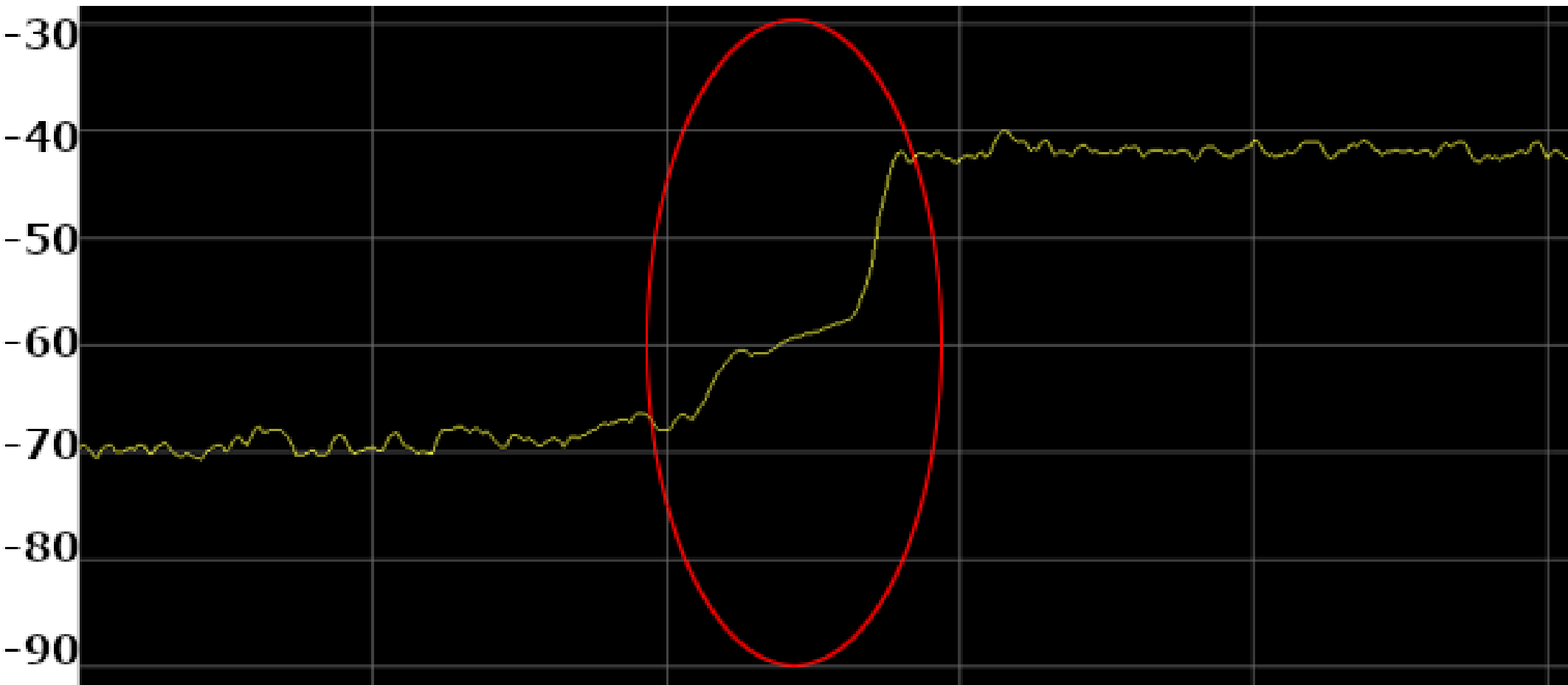}
      \caption{Transmit gain 79 with moderate distortion.}
    \end{subfigure}
    \begin{subfigure}{0.47\textwidth}
      \includegraphics[width=\textwidth]{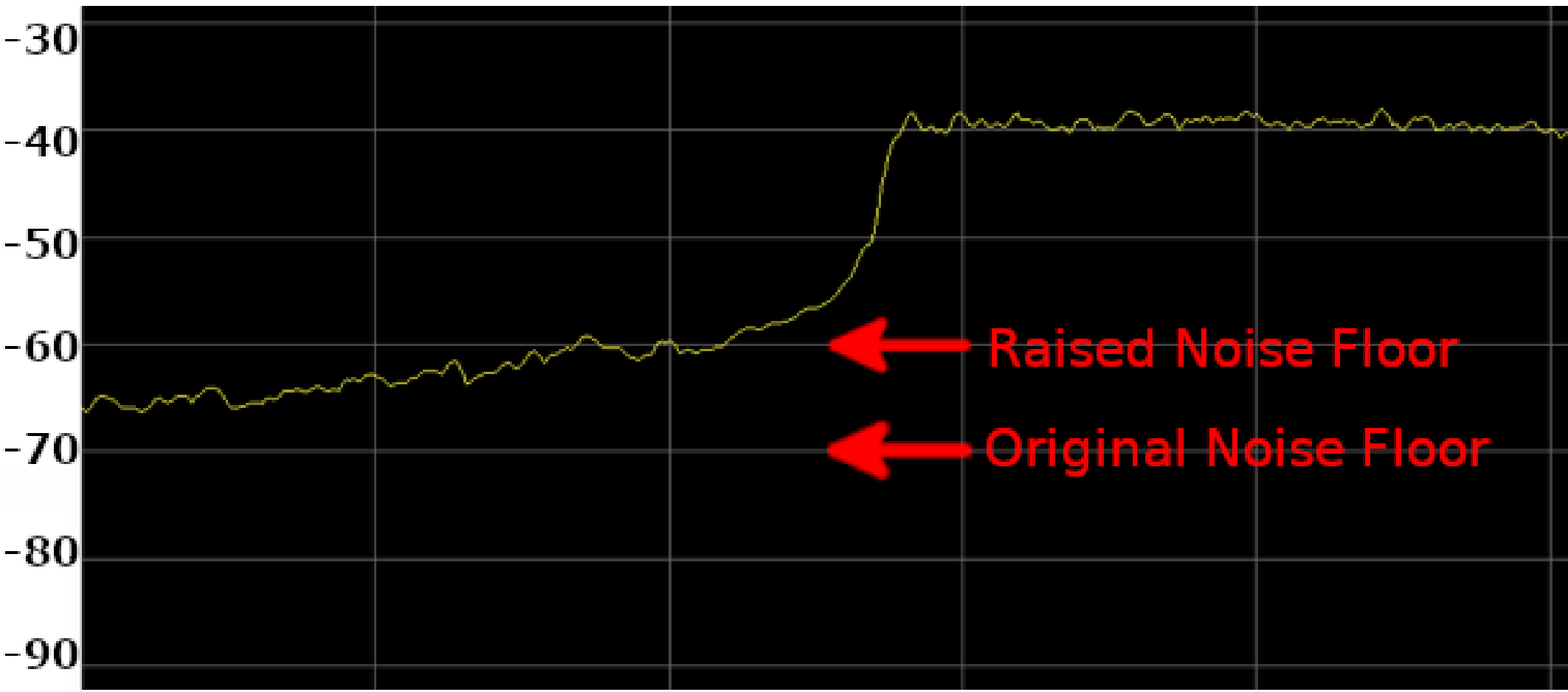}
      \caption{Transmit gain 82 with large distortion and raised noise floor.}
    \end{subfigure}
    \caption{Progression of distortion increasing with transmit gain at 20 MHz.}
\end{figure}

\begin{figure}
\vspace{-7mm}
\begin{center}
\includegraphics[width=0.5\textwidth]{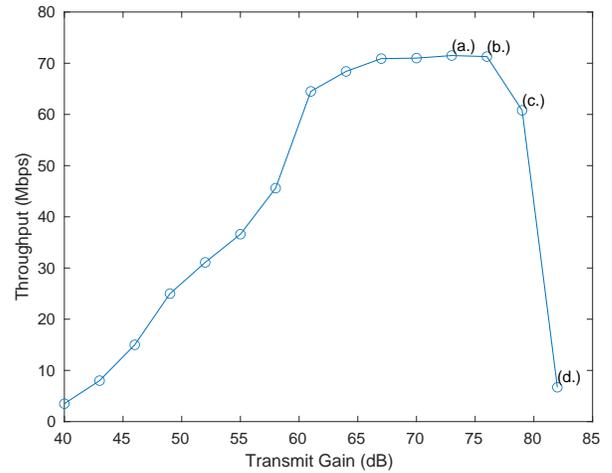}
\end{center}
\vspace{-5mm}
\caption{Throughput of increasing transmit gains at 20 MHz. The (a.)-(d.) points are associated with Figs. 1(a)-1(d).}
\label{fig:distortionthroughput}\vspace{-5mm}
\end{figure}


\textbf{Appearance of the Waveform: }
Monitoring the spectral appearance of any transmitted signals is important both for optimizing performance and keeping the transmitted signals within their legal bounds. This can be done by using dedicated spectrum analyzing equipment or SDRs tuned to receive on the transmitted frequencies, possible spurious transmission in adjacent bands, harmonics, and so forth. When observing the signal, it is therefore important that the device shows a larger bandwidth than the transmitted signal and that it has a low noise figure. 
A larger bandwidth allows the user to monitor if the signal is using the proper amount of bandwidth expected or if there is distortion outside the expected bandwidth. As the SDR gain increases, the internal power amplifier (PA) may become saturated and show "shoulders" on the sides of the signal. These are caused by intermodulation distortion that happens when two or more carriers/subcarriers drive an RF device into it nonlinear operating region. Continued gain increases will increase the power and width of this distortion and may eventually cause the surrounding noise floor to increase noticeably. The noise floor during transmission should be compared to the noise floor when no transmission is occurring. Any differences in power may indicate distortion caused by the transmitter and may impact the radio performance or interfere with other devices. Figure 1 shows the progression of distortion as the USRP transmit gain is increased. Figure \ref{fig:distortionthroughput} shows the throughput at each transmit gain including labels matching their spectrum in Figure 1. In Figure 1(a) there is a small amount of distortion at the edges of the signal using a transmit (Tx) gain of 73 for a 20 MHz LTE downlink signal. This corresponds with the maximum throughput in Figure \ref{fig:distortionthroughput}, which is close to the theoretical maximum of about 75 Mbps. Figure 1(b) shows out of band distortion through the appearance of shoulders and a small decrease in throughput in Figure 2. This trend continues at a Tx gain of 79 when the throughput decreases by more than 10\%. Finally in Figure 1(d) we see that the noise floor raise significantly along with a significant decrease in throughput. Based on these results, our typical experiments use a Tx gain of 73 for maximum output with minimal distortion and interference outside of our chosen band.
These results were obtained with a B205mini-i USRP, a 15 W PA, and 4400 MHz cut-off low pass filter in the transmit chain. Results were consistent across different bands.

\textbf{Comparison of SDR Gains:}
There are several options available for SDRs based on the required application. All SDRs will have the ability to adjust the internal PA for the Tx and receive (Rx) chains. However, the gain values used for each device will not have the same output as different models using those values. An example of this is comparing the USRP X310 and B205mini-i. The B205 has a maximum gain of 89.75 whereas the X310 maximum is 31.5. It can not be inferred that the B205 is capable of outputting more power due to a larger range of gain values. A comparison should be done across devices to determine what gains are similar across devices and are safe to use without causing distortion.

\textbf{Throughput: }
A common benchmark for determining the performance of an experiment is the ability for data to be communicated over the channel. Pinging devices over the channel is a low bandwidth test to determine whether or not the devices are able to perform basic communication over a channel and the stability of the channel based on the latency values provided while pinging. Additionally, a process such as iperf may be used to stress test the channel by utilizing its full capacity. This will indicate the maximum amount of data the channel is capable of handling and further tests stability based on the consistency of the obtained values. 

\textbf{Connection Distance: }
Connection distance serves as a good benchmark when range/coverage area is of interest. This benchmark is most relevant when used alongside throughput benchmarks. It helps for optimizing gains, frequency usage, and effects of antenna orientation. When using a UAV, various heights can also be taken into account to determine how it impacts the connection over vertical and horizontal distances.

\subsection{Methodology}

The process by which each benchmark is performed may depend on the specific use case. Our methodology for developing and optimizing LTE handover experiments with open-source SDRs is discussed in continuation.

\textbf{Appearance of the Waveform: }
The waveform appearance  benchmark requires a real-time or near real-time spectrum analysis instrument. 
Monitoring should begin before any other devices are connected to measure the baseline with no experimental signals. 
Once the experiment begins, the signal should be monitored for any unexpected behavior such as distortion in the form of shoulders or a raised noise floor. If either type of distortion is present, it is likely that saturation is occurring and causing unwanted operation. The Tx gain of the device should be reduced before repeating the experiment. Although the signal produced by an SDR may be clean and non-distorted, it is still possible for the signal to become distorted when additional external components are added such as PAs. Performing the benchmark on an SDR expected to receive signals will also be useful ensuring an appropriate Rx gain is used to prevent distortion caused at the receiver. Receivers are nonlinear and distortion can cause severe link degradation \cite{dsouza2020symbol}. This benchmark should be performed any time new SDRs or RF components are used for an experiment. When performing handover experiments, performing this benchmark is important across both eNBs and the UE. Both eNBs and the UE should transmit clean signals to avoid interfering with the other nodes, which could make the process of determining the proper time to handover more difficult.

\textbf{Comparison of SDR Gains: } 
Comparing SDR gains is most useful when multiple SDR types are expected to be used or a transition of types for a node is expected. When performing an experiment, this benchmark can also be useful for ensuring SDRs of the same type are producing the same output when given the same input. This benchmark will also require a spectrum analyzer or an SDR for monitoring. To compare either Tx or Rx gains of a set of devices, the devices should be co-located and using the same front end components. To compare Tx gains, an appropriate Tx gain should be selected for the devices and operated one at a time with the result recorded on the monitoring device. By doing so, the difference in power can be determined on the monitoring device and the Tx gains can be changed to determine the pair that produces most similar output. Similarly, Rx gains can be compared by flipping the roles from the Tx gain comparison. It is likely for handovers to be performed over different SDR devices based on the fixed vs. mobile nature of different nodes. Therefore, proper gains should be selected on each device for optimal output from each device.

\textbf{Throughput:}
Ping and throughput benchmarks are ideal for testing the quality and strength of a signal once a communication channel is formed. At least two nodes that are able to share data are required for this benchmark. Ping should be used initially to confirm that each node is able to communicate with the interface at every other node it should have access to. This will also give an initial indicator to the stability of the channel based on the variance of the latency reported by the ping process. To test the maximum capacity and stability of the channel, iperf is used to utilize all the resources available. Iperf will attempt to pass as much data as possible through the channel. This benchmark can be done in simultaneously or individually in multiple directions to test several channels and their hardware. The stability of a channel is also indicated based on the consistency of the throughput measured. Ping and throughput tests are ideal for testing the effectiveness of handovers. A ping test will help determine the amount of latency introduced, and a throughput test will determine how much data is lost during the handover.

\textbf{Connection Distance:}
Connection distance is useful for determining the coverage area of a signal. Range tests are best done when the UE is mobile and can easily be repositioned for several data points. A UAV can easily move to various points and provides height as an added measurement. The goal of this benchmark is to determine the range at which a connection between the eNB and UE can be held. When combined with a throughput test, data can be gathered regarding latency and quality of connections at various distances and heights. Distance is a vital measurement for setting up handover experiments. If BSs overlap coverage area's with too strong of power, this will interfere with both performance and the ability to produce a proper handover. On the other hand, if there is too much distance, the connection may drop entirely between BSs or ping pong back and forth without performing a handover.

\subsection{Hardware Selection}

Hardware selection can be an iterative process involving selecting SDRs and their RF front ends based on the experiments that are expected to be operated and the results produced from each iteration of an experiment. Reference \cite{ACM_SDR} breaks down several available SDR options based on their features and scenarios where certain SDRs will be more effective than others. Similar considerations should be taken when determining the appropriate components for a front end. Benchmarking may indicate a requirement to use PAs, for example, to increase coverage area. However, this may increase distortion and weight while requiring another power source.

\begin{table*}[]
\caption{Open-source software core networks and their supported handover types for LTE.}
\label{tab:HandoverComparison}
\begin{tabular}{p{1cm}|p{5.1cm}|p{5.0cm}|p{5.0cm}|}
\cline{2-4}
\multirow{3}{*}{} & \multicolumn{3}{c|}{\textbf{\small Handover}} \\ \cline{2-4} 
                  & X2     & S1    & Intra-eNB    \\ \hline
\multicolumn{1}{|c|}{\small srsEPC} &
  \small Not supported. SrsRAN is expected to introduce X2 handover in future releases. &
  \small Not Supported. &
  \small Supported. The handover occurs between one sector to another. \\ \hline
\multicolumn{1}{|c|}{\small open5GS} &
  \small Supported. &
  \small Supported. The handover occurs over S1-interface with the help of MME. &
  \small Supported. \\ \hline
\end{tabular}
\end{table*}
The srsRAN supports two different handover mechanisms, S1 and intra-eNB. The CN plays an important role in executing the handover process. 
Table~\ref{tab:HandoverComparison} compares two CNs that are compatible with srsRAN in terms of their support for the different LTE handover types. 
It is worth mentioning that srsEPC supports 
handover between two radio chain of the same radio device/SDR, such as a B210 USRP that has two Tx/Rx chains. 
Because of practicality, we integrate the open5GS CN with srsENB in order to support true handover functionality. 

\section{Advanced Wireless Using ZMQ}
\label{sec:simulation}

ZMQ is a messaging library that allows for the simulation of srsRAN communications. It is used as an RF driver by sending IQ samples using inter-process communications (IPC) or the transmission control protocol (TCP). 

\subsection{Why ZMQ?}

SrsRAN is typically used with radio hardware, such as USRPs or LimeSDR. ZMQ gives users access to srsRAN without the high initial investment costs associated with radio experiments. In the instance that SDRs are available, ZMQ allows for more rapid development of experiments in srsRAN for testing prior to real world experiments. This can be done as a simple experiment from srsRAN node to another srsRAN node or made more complex by using custom written programs using ZMQ or GNU Radio via the ZMQ blocks. GNU Radio gives access to several ZMQ patterns such as request/reply, push/pull, and others. By default, srsRAN uses the request/reply pattern with ZMQ. When used with a custom program or GNU Radio, signal processing can be performed to create advanced experiments such as S1 handover or multiple UEs.

\subsection{ZMQ with GNU Radio}

When running srsRAN with ZMQ and no external programs, experiment complexity is limited. The eNB will connect to the UE over two ZMQ ports, one for downlink and one for uplink. This creates a standard single input, single output (SISO) connection where the user can exchange data through programs such as ping or iperf. To increase complexity, the ZMQ request and reply blocks can be used to connect to Tx and Rx ports on the eNB and UE. This pushes the data from srsRAN through GNU Radio instead of exclusively through srsRAN. Once the data is passed to GNU Radio, the Tx data from each node can be directly passed to the Rx port of each node to recreate the initial SISO experiment.


\subsection{Using Open5GS}
In order to perform S1 handover, srsEPC can not be used as it does not yet support this feature. Instead, srsRAN recommends using Open5GS. Open5GS is another CN solution that supports both 4G and 5G. Once installed, parameters between the software stacks will need to be synced such as IP addresses and the public land mobile network values. Additionally, the UE will need to use the milenage algorithm and be added as a subscriber to the Open5GS database. This can be done via a Web user interface (UI) supported by Open5GS.

\subsection{S1 Handover Setup}
\begin{figure}
\vspace{-3mm}
\begin{center}
\includegraphics[width=0.48\textwidth]{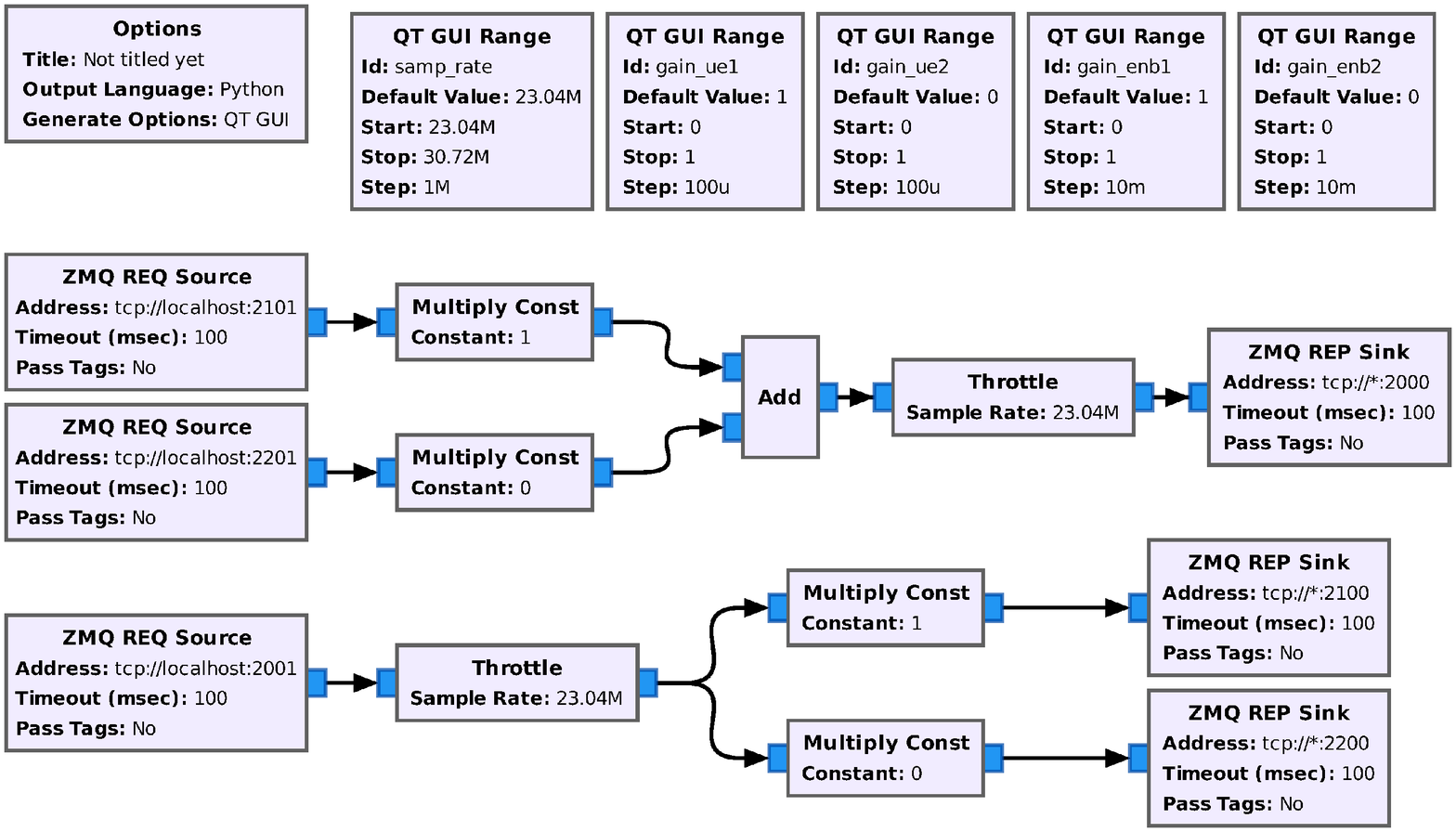}
\end{center}
\vspace{-5mm}
\caption{GNU Radio flowgraph using srsRAN and ZMQ for S1 handover.}
\label{fig:ZMQ_Handover}\vspace{-5mm}
\end{figure}

S1 handover can be performed by using two eNBs and a single UE. Figure \ref{fig:ZMQ_Handover} shows the GNU Radio flowgraph that can be used for S1 handover. Once the ports are set up on the srsRAN nodes, the experiment can be started as a SISO experiment with one channel pair disabled via the multiply constant blocks. 
The experiment starts with a single channel pair enabled by setting the downlink and uplink channels associated with the first eNB to 1 and all other to 0. All multiply constant blocks are controlled by QT graphical UI (GUI) Range blocks to allow varying values to be set throughout the experiment.
To perform the handover, the multiply constant pairs can be increased for the eNB that the UE will move towards and decreased for the eNB that the UE is moving away from. This will simulate movement between the BSs. When the signal of the second eNB becomes sufficiently stronger than the initial eNB, the initial eNB will initiate the S1 handover. If successful, the second eNB will acknowledge this request and the UE will note that a successful handover took place. 

\subsection{Using Throttle}
Due to the simulation nature of using ZMQ with srsRAN, the channels between the eNB and UE will have higher capacity than when used with real SDR hardware that are limited by sampling rates. In order to increase the realistic nature of the simulation, a throttle is added to the channels in the GNU Radio flowgraph to restrict sample rates. In order to recreate a real srsRAN experiment as closely as possible, a throttle using a 3/4 sample rate with respect to the number of resource blocks is used due to its use with many SDRs in srsRAN.

\subsection{S1 Handover Results}
\begin{figure}
\vspace{-3mm}
\begin{center}
\includegraphics[width=0.45\textwidth]{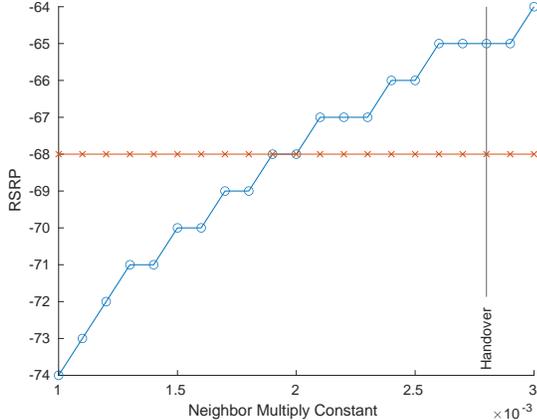}
\end{center}
\vspace{-5mm}
\caption{S1 handover using ZMQ from local BS to neighbor.}
\label{fig:ZMQ_HO_Result1}\vspace{-5mm}
\end{figure}

\begin{figure}
\vspace{-1mm}
\begin{center}
\includegraphics[width=0.45\textwidth]{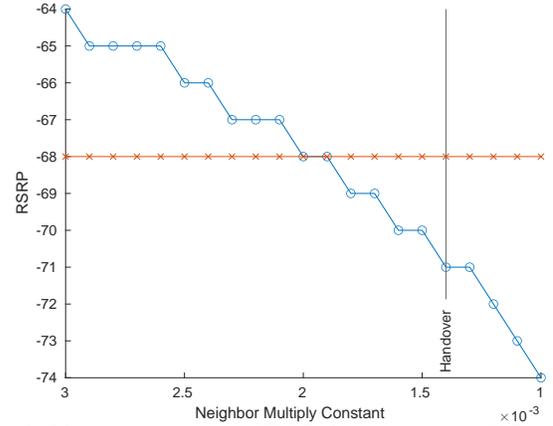}
\end{center}
\vspace{-5mm}
\caption{S1 handover using ZMQ from neighbor BS to local.}
\label{fig:ZMQ_HO_Result2}
\vspace{-5mm}
\end{figure}

When performing the simplest case of S1 handover using ZMQ, results are generally consistent based on the input parameters. Ping and iperf tests will have some level of variability around the expected scenario, Using the throttle block to restrict to common sample rates will restrict the variability around the expected value at that sample rate. Other values such as reference signal received power (RSRP) and signal-to-noise ration (SNR) will be quite consistent based on the values of the multiply constant block. When a constant value of 1 is used, the signal will be very strong with RSRP values stronger than -20 and SNR values over 100. The constant must be decreased to a much smaller value in order to get more realistic values. In our experiments, the local BS is kept at a constant RSRP of -68 to determine at what when srsENB would initiate the S1 handover. The neighbor BS is given a multiply constant value simulating a weaker signal and increased over time. It is found that the S1 handover is initiated when the RSRP of the local BS is 3 less than that of the neighbor station. Figure \ref{fig:ZMQ_HO_Result1} shows the neighbor station's signal increasing until the handover occurs. In Figure \ref{fig:ZMQ_HO_Result2}, the neighbor station has a stronger signal and is initially connected to the user. The signal strength is decreased until the handover occurs when the RSRP of the local station is 3 greater than the neighbor station, the same value as the previous case.

\section{Advanced Wireless Using UAVs}
\label{sec:outdoor}
The transition from simulation to actual field testing should be incremental to allow for early detection and correction of problems. The first step might involve the SDRs connected to each other with coaxial cable and no other RF components other than attenuators. This enables the confirmation of SDR settings and other software configuration. The next step might be to replace the cables with antennas. Range is very limited in this setting but whatever interference is present in the environment will effect the experiment. Next, the transition to the full RF chain that will be used on the UAV and BS is accomplished in the lab. The transmit chain includes a low pass filter to remove harmonic artifacts, a power amplifier (15 W for BS and 1 W for UAV), and the antenna. The receive chain includes the antenna, a band pass filter and low noise amplifier. The use of attenuators after the PAs and/or a reduction in gain settings might be needed to avoid saturating the receive link at these close distances. 

\subsection{Testing Outdoors with UAVs}

Other than the obvious difficulties of weather and set up/tear down times, working in the field with UAVs has some other issues. The RF link in the open field will be different than in the enclosed laboratory, so one needs to be prepared for new difficulties. There is no automatic gain control function in srsLTE, so manual adjustment will be necessary. Lower gains work better at closer distances and high gains at larger distances. Power management on the UAV is critical for flight time. We elected to use a separate battery for the payload which is only used when airborne. Otherwise, a corded power supply is used while setting up and troubleshooting. Switching between the two can be done without interrupting the power supply to the computer. It is beneficial to select all amplifiers in such a way to share a common power supply for simplicity and weight savings on the UAV. 

If WiFi is needed, one should consider the placement of the computer if the WiFi antennas are internal. Upside down placement on the outside-bottom of the payload container ensures less attenuation to provide longer range. 
The radio antennas need to be separated from each other and from the RC antennas to avoid self and cross-interference. The positioning of the antennas should be done also accounting for how the banking of the UAV may potentially block the line of sight path between the ground node and aerial node antennas. 
\vspace{-3mm}
\subsection{Backhaul}
The radio access network (RAN) connects to the CN through the backhaul. The backhaul is thus needed for accessing CN services, such as user authentication and mobility management, but also to access external networks and application servers. 
Wired (fiber) backhaul is typical for commercial cellular networks, but wireless backhaul via a directional RF link is also deployed. 

A cellular network implemented with SDRs has multiple options: (1) IPC when the RAN and CN are implemented on the same computer, (2) wired Ethernet connections over copper or fiber, or (3) wireless backhaul, e.g. through a dedicated RF link or via a WiFi
access point (AP). 
IPC is only possible in a Cloud RAN, where the baseband processing of the BSs and the CN can be collocated, but then the fronthaul needs to be provided via copper/fiber or via radio \cite{MICRO12}. 
Using 3GPP terminology, the backhaul is implemented over the S1 interface. In practice there is one IP connection between each BS and the CN.
If implemented over copper or fiber or a dedicated RF link, the data rate is easily sustained and the latency will be minimal. 
If implemented via an WiFi AP, the latency will increase and be less deterministic because of the WiFi channel access. The bandwidth can become a problem for long wireless links or challenging channel conditions. 
An experimental WiFi network is the most flexible and non-proprietary solution 
where there is no wired network access. We recommend directional antennas between the WiFi stations and AP for range, and MU-MIMO (available with IEEE 802.11ax or WiFi6) to establish multiple high-speed links serving two or more BSs. 
Measurements have shown that using WiFi for implementing the backhaul leads to larger latency and, more importantly, large fluctuation in latency with an order of magnitude difference (10s to 100s of ms). The throughout can be a limitation for long distances between the AP and the WiFi stations/BSs. Additionally, a 2.4 GHz WiFi signal can be a source of interference to the 
radio experiment if using a nearby frequency. WiFi6 can be operated at 5 GHz.

\begin{figure}
\begin{center}
\includegraphics[width=0.45\textwidth]{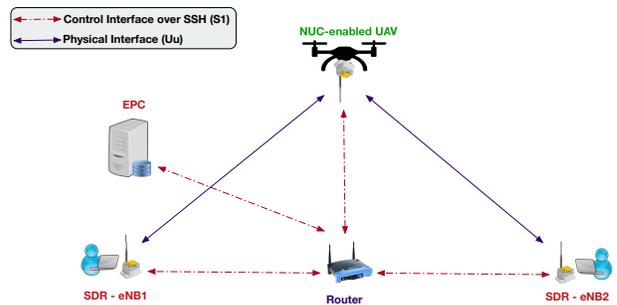}
\end{center}
\vspace{-3mm}
\caption{WiFi-assisted backhaul for outdoor SDR testbed.}
\label{fig:backhaul}
\end{figure}

Figure \ref{fig:backhaul} shows the overall backhaul of our system. A router is centrally located with respect to the other nodes. The LTE CN---the evolved packet core (EPC)---and both eNBs are connected to each other using WiFi for a wireless S1 interface. In addition to being used for the S1 interface, the WiFi connection is also used to ssh into the EPC, eNBs, and UE. Each eNB uses WiFi for the backhaul connection and SDRs for the cellular connection to the UE.

\begin{figure}
\vspace{-5mm}
\begin{center}
\includegraphics[width=0.48\textwidth]{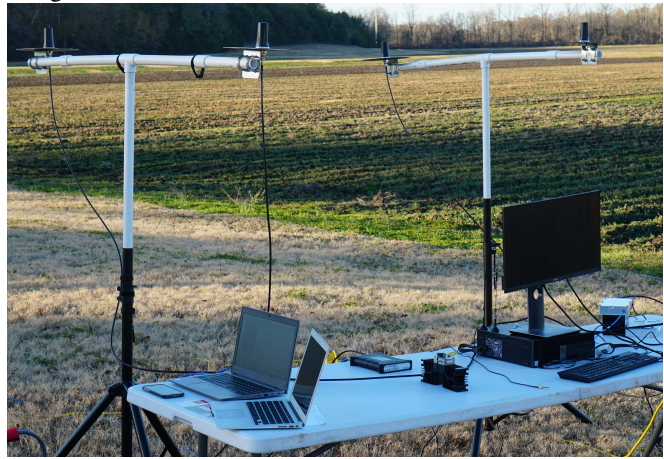}
\end{center}
\vspace{-3mm}
\caption{SDR BS with RF, computer, and generator.}
\label{fig:BS}
\end{figure}

\begin{figure}
\vspace{-5mm}
\begin{center}
\includegraphics[width=0.48\textwidth]{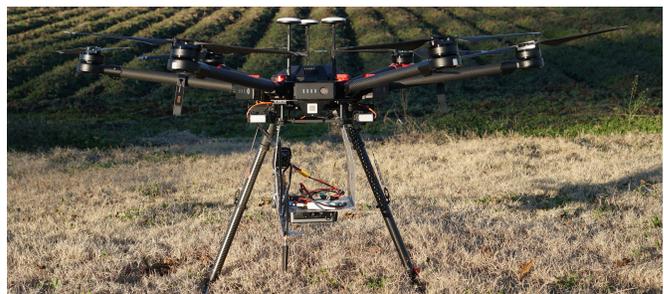}
\end{center}
\vspace{-3mm}
\caption{UAV with SDR payload.}
\label{fig:BSwUAV}
\end{figure}

\subsection{Preparing Large Scale S1 Handover}

Performing S1 handover in an laboratory can be more difficult than doing so in an open area outdoors. In a laboratory, the SDRs will be exposed to more multipath signals due to the presence of walls, tables, and other objects in the room. Additionally, without a large distance between BSs, the SDRs will rely on their internal amplifiers allowing for very short ranges at higher frequencies. In order to do handover, the coverage of each BS must overlap which is a regular phenomenon in practical scenarios at the cell edge and allowing smooth handover with slight disruption in transmission.
If the BSs do not overlap enough, this will cause 
the UE to disconnect and reconnect to the other BS without performing a handover. In an open-field rural setting, multipath reflections may be less pronounced. 
Additionally, using external PAs is viable since larger distances can be created between the radios to avoid saturation. This creates larger coverage areas allowing for larger overlap and transition areas where the handover can occur. Figures 7 and 8 show the ground and aerial node prototypes for the outdoor experiments.

\subsection{S1 Handover Results}
\begin{figure}[t]
\begin{center}
\includegraphics[width=0.47\textwidth]{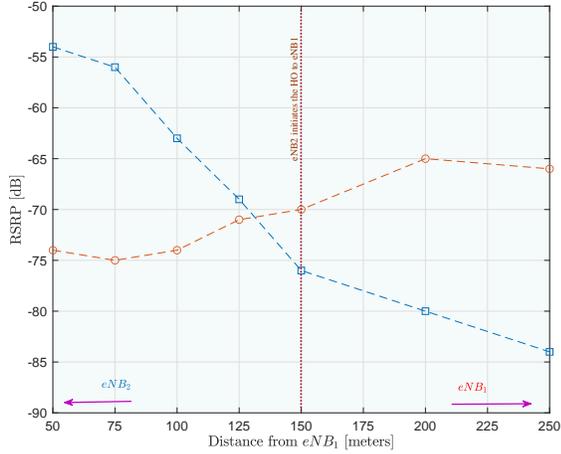}
\end{center}
\vspace{-5mm}
\caption{S1 handover from eNB2 to eNB1}
\label{fig:Outdoor2}
\end{figure}

Experiments were run with the EPC separated from both eNBs. The eNBs were approximately 525 m apart from each other. The UAV would hover at the desired height, 45 m in our experiments, and measurements would be recorded based on the traces reported by the srsUE. The eNBs traces indicate the initiation and reception of S1 messages from the EPC. The UE trace reports the RSRP to the eNB that it is currently connected to and the neighbor eNB. These values are used to determine when the handover should occur. The UAV flies on a trajectory between the two 
eNBs. Figure \ref{fig:Outdoor2} shows the UE connected to eNB2 and flying towards eNB1. As the UAV flies away from eNB2 and towards eNB1, the RSRP values become closer to each until overlapping around 125 m. The handover occurs after the RSRP favors the neighboring cell, similar to the results observed during the handover evaluation with ZMQ presented in Figure 5. It happens at the 150 m mark here (Figure 9). In the opposite direction (Figure 10), the 
handover occurs around the 115 m mark, after the reported RSRP values start to diverge.


\begin{figure}[t]
\begin{center}
\includegraphics[width=0.47\textwidth]{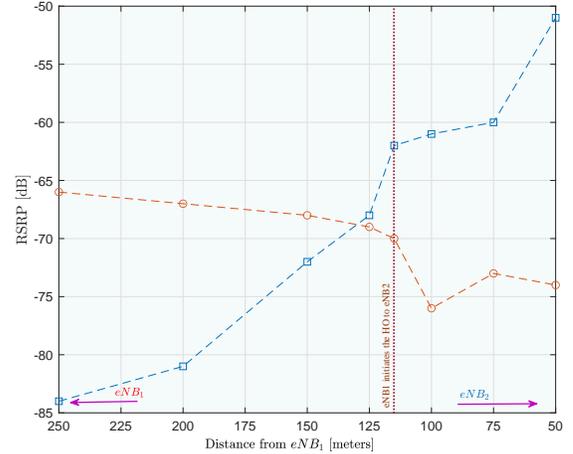}
\end{center}
\vspace{-5mm}
\caption{S1 handover from eNB1 to eNB2}
\label{fig:Outdoor3}
\end{figure}




\section{Conclusions}
\label{sec:conclusions}
This paper has presented our approach for performing real handover experiments using SDRs and UAVs. We discuss several benchmarks that can be used for both optimizing and qualifying the effectiveness of an experiment's performance. A simple handover simulation is designed using srsRAN, ZMQ, and GNU Radio for testing the expected behavior of the handover. We further discuss how to set up a handover experiment in the field using UAVs with results showing the real world behavior of the handover. Future steps include performing similar experiments using UAVs for advanced cellular functions such as MIMO and carrier aggregation.

\section*{Acknowledgment}
This work was supported in part by NSF award CNS-1939334. 


\bibliographystyle{IEEEtran}
\bibliography{imsi,vuk,CAREERisma,isma}

\end{document}